\documentclass[a4paper]{jpconf}
\usepackage{graphicx}
\begin{document}
\title{Does quantum mechanics tell an atomistic spacetime?}

\author{Hans-Thomas Elze}

\address{Dipartimento di Fisica ``Enrico Fermi'',  
        Largo Pontecorvo 3, I-56127 Pisa, Italia }

\ead{elze@df.unipi.it}

\begin{abstract}
The canonical answer to the question posed is ``Yes." -- tacitly assuming that quantum theory and the concept of spacetime are to be unified by `quantizing' a theory of gravitation. Yet, instead, one may ponder: Could quantum mechanics arise as a coarse-grained reflection of the atomistic nature of spacetime? -- We speculate that this may indeed be the case. We recall the similarity between evolution of classical and quantum mechanical ensembles, according to Liouville and 
von\,Neumann equation, respectively. The classical and quantum mechanical equations are indistinguishable for objects which are free or subject to spatially constant but possibly time dependent, or harmonic forces, if represented appropriately. This result suggests a way to incorporate anharmonic interactions, including fluctuations which are tentatively related to the underlying discreteness of spacetime. Being linear and local at the quantum mechanical level, the model offers a decoherence and natural localization mechanism. However, the relation to primordial deterministic degrees of freedom is nonlocal.
\end{abstract}
%%%%%%%%%%%%%%%%%%%%%%%%%%%%%%%%%%%%%%%%%%%%%%%%%%%%%%%%%%%%%%%%%%%%%%%   
\section{Quantum features, information loss and spacetime discreteness}

Classical ensemble theory with its 
Liouville equation implies the absence of a stable ground state, when rewritten 
as a Hilbert space theory with an analogue of the Schr\"odinger 
equation~\cite{Koopman,vN,tHooft06a}. 
Correspondingly, the analogue of the von\,Neumann equation obtains an 
unusual superoperator which couples the Hilbert space and its 
dual~\cite{Elze09}, cf. Section~2 below. 

These features have presented serious obstacles to a number of recent   
attempts to understand quantum mechanics as an 
emergent phenomenon~\cite{tHooft06b,Elze05,Blasone05,Adler,Smolin,Vitiello01}, 
see also Refs.~\cite{Scardigli09,Isidro08,Wetterich08}. 

Such studies are strongly 
motivated by issues surrounding quantum gravity, in particular, 
the fact that quantum theory is hardly compatible with the symmetry requirements  
of general relativity or of other theories describing gravity and spacetime, which  
are based on general coordinate invariance. 
Furthermore, despite the great successes in explaining   
statistical aspects of experiments, the intrinsically indeterministic features 
of quantum mechanics remain problematic. This is seen, for example, in the unresolved   
measurement problem, with related issues of wave function collapse or objective 
reduction~\cite{GianCarlo09,BassiGhirardiRev}, and in the   
quantum mechanics of the universe, considered as a whole, 
which might be self-contradictory.   

Yet there is a proof of existence of deterministic models for quantum mechanical 
objects, in which dissipation, i.e., a fundamental information loss 
mechanism, has been an essential ingredient~\cite{tHooft07,Elze08}.    

With these findings in mind, we presently discuss aspects 
of a dynamical transition {\it from} classical {\it to} quantum behaviour,   
assuming that: {\it Quantum features originate from a dissipative process which affects all 
physical objects. Dynamics and statistical interpretation of quantum states 
(Born rule) originate from deterministic rules, such as embodied in classical mechanics, 
in an ensemble theory.}   

We speculate that the {\it atomistic} structure of spacetime itself   
is responsible for effects which are attributed to quantum mechanics, 
typically operating at length scales very much larger than the Planck length. 
Here ``atomistic'' refers to a discrete 
set of elements with the presence, or absence, of a certain order relation  
between any two elements. Furthermore, this set, in particular 
the number of its elements changes dynamically, 
possibly establishing new order relations, or erasing old ones. 
In this way, ``time happens''.   

Similar ideas about the nature of spacetime have been formulated every now and then 
throughout the history of natural philosophy~\cite{Jammer,SorkinRev}. 
However, only recently such general scenario 
has been elaborated in more detail in the theory of {\it causal sets}. 
In mathematical terms, these are locally finite ordered sets. Their evolution  
by sequential growth through random (``sprinkling'') appearance of new set 
elements together with their order relations has been studied~\cite{SorkinRev,Dowker,RideoutSorkin1}. 

In the absence of an equally elaborate theory of matter in relation to such atomistic 
structure of spacetime, we can nevertheless state the following, concerning the 
situation of a typical object. Consider an electron, for example, an object that 
to highest known precision behaves according to the laws of quantum mechanics. 
We may term its ``situation'' a hypothetical complete set of its properties 
that are accessible by experiments. 

Now, first of all, there are interactions between 
such object and its environment (the ``rest of the universe''), 
last not least gravitational ones. Besides more familiar aspects, however, 
there must exist a continual 
{\it information loss} about its situation, since the  
atomistic spacetime beneath evolves. With respect to the latter, 
quantum theory at present deals with very coarse-grained phenomena, when describing 
the dynamics of matter. 
Loosely speaking, in order to fully characterize an electron, the evolution of its causal relations with a continually changing number of spacetime ``atoms'' has to come into play. 
Consequently, in a coarse-grained picture where this is not explicitly taken into account, 
information about an object degrades, due to the evolving -- always and 
everywhere present -- environment which is spacetime.  

Secondly, however, common objects are characterized by a certain persistence, 
which makes them identifiable in experiments.  
Therefore, the information loss must be a delicate one. 
It must be compatible with the {\it conservation of probability}, which is a basic 
tenet of a reasonable ensemble theory.

Contrary to measurement processes in quantum mechanics, where information is transferred from 
microscopic to macroscopic objects,~\footnote{In a way which is not understood in all 
its aspects in 
theory. For recent discussions, which critically review related ideas, 
see Refs.~\cite{GianCarlo09,BassiGhirardiRev}.}  
we propose here that matter degrees of freedom are  
continually ``measured'' by the evolution of spacetime.      

These heuristic considerations lead us to modify    
the classical ensemble theory in important ways. We will  
incorporate dissipation into the Liouville equation, however,  
in such a way that probability conservation remains intact.   
This provides us with a glimpse of the mechanism that might be 
responsible for turning the deterministic evolution of classical objects, 
described here by an ensemble theory, into the Schr\"odinger evolution of quantum objects.  

%%%%%%%%%%%%%%%%%%%%%%%%%%%%%%%%%%%%%%%%%%%%%%%%%%%%%%%%%%%%%%%%%%%%
\section{The Liouville equation in terms of Hilbert space operators}

It will be shown here how the {\it classical} Liouville equation 
can be written in the form of the {\it quantum mechanical} von\,Neumann equation, 
generally, incorporating a characteristic extra term. 

We consider objects with a single continuous degree of freedom, for 
simplicity.~\footnote{The following is easily repeated for few-body systems or fields.}  
To begin with, we consider conservative forces, such that the equations 
of motion are determined by the generic Hamiltonian function:  
\begin{equation}\label{HamiltonianF} 
H(x,p):=\frac{1}{2}p^2+v(x) 
\;\;, \end{equation} 
depending on generalized coordinate $x$ and momentum $p$, and where  
$v$ denotes the {\it true potential}. In Section~3, we 
will come back to the notion of the true potential and distinguish it 
from a related {\it coarse-grained potential} $V$. 

An ensemble of such objects, for example, 
following trajectories with different initial conditions, is described by 
a distribution function $f$ in phase space, i.e., by the probability 
$f(x,p;t)\mbox{d}x\mbox{d}p$ to find a member of the ensemble in an infinitesimal 
volume at point $(x,p)$. This distribution evolves according to the Liouville equation: 
\begin{equation}\label{LiouvilleEq} 
-\partial_tf=\frac{\partial H}{\partial p}\cdot\frac{\partial f}{\partial_x}
-\frac{\partial H}{\partial x}\cdot\frac{\partial f}{\partial_p}
=\big\{ p\partial_x-v'(x)\partial_p\big\}f 
\;\;, \end{equation} 
with $v'(x):=\mbox{d}v(x)/\mbox{d}x$. -- We recall that the relative minus sign 
in the Poisson bracket, or between terms here,   
reflects a symplectic phase space symmetry. This will translate into the familiar 
commutator structure in Eq.\,(\ref{Schroed}).   

A Fourier transformation, $f(x,p;t)=\int\mbox{d}y\;\mbox{e}^{-ipy}f(x,y;t)$, 
replaces the Liouville equation by: 
\begin{equation}\label{LFourier}  
i\partial_tf=\big\{ -\partial_y\partial_x+yv'(x)\big\}f 
\;\;, \end{equation} 
without changing the notation for the distribution function,  
whenever changing variables.  
Thus, momentum is eliminated and a {\it doubled number of coordinates} results. 
Finally, 
with the transformation: 
\begin{equation}\label{coordtrans} 
Q:=x+y/2\;\;,\;\;\;q:=x-y/2  
\;\;, \end{equation} 
we obtain the Liouville equation in the form: 
\begin{eqnarray}\label{Schroed} 
i\partial_tf(Q,q;t)&=&\Big\{ \hat H_Q-\hat H_q+{\cal E}(Q,q)\Big\}f(Q,q;t)
\;\;, \\ [1ex] \label{HX} 
\hat H_\chi &:=&-\frac{1}{2}\partial_\chi ^{\;2}+v(\chi )\;\;, 
\;\;\;\mbox{for}\;\;\chi =Q,q 
\;\;, \\ [1ex] \label{I} 
{\cal E}(Q,q)&:=&(Q-q)v'(\frac{Q+q}{2})
-v(Q)+v(q)\;=\;-{\cal E}(q,Q)
\;\;. \end{eqnarray}  
Several comments are in order here: 
\begin{itemize} 
\item The present reformulation of classical dynamics in phase space can be carried out  
rather independently of the number of degrees of freedom and is applicable   
to matrix or 
Grassmann valued variables as well; see, for example, Refs.~\cite{Elze05,Elze07}. 
Gauge theories or, generally, theories with constraints 
have to be examined carefully.  \\ \noindent  
\item Most importantly, the Eq.\,(\ref{Schroed}) closely resembles the 
{\it von\,Neumann equation} for a density operator $\hat f(t)$, 
considering $f(Q,q;t)$ as its matrix elements. 
We automatically recover the Hamiltonian operator $\hat H$  
related to the Hamiltonian function, Eq.\,(\ref{HamiltonianF}),  
as in quantum theory. 
However, an essential difference consists   
in the interaction $\hat {\cal E}$ between 
{\it bra-} and {\it ket-} states. 
The Hilbert space and its dual here are coupled by 
a {\it superoperator}.~\footnote{This superoperator is of a very    
specific form, which leads to the antisymmetry in 
Eq.\,(\ref{I}). It differs from a Lindblad superoperator, often  
obtained as a symmetric double commutator structure, 
in the case of open quantum mechanical 
systems~\cite{Diosi}; this is seen, for example, in our Eq.\,(\ref{vNLM}) below.} \\ \noindent     
\item Alternatively, the Eq.\,(\ref{Schroed}) might be read as the     
{\it Schr\"odinger equation} for two identical (sets of) degrees of freedom. 
However, their respective Hamilton operators, $\hat H_{Q,q}$, contribute with opposite sign, 
which must be traced to the classical symplectic symmetry. 
Since their interaction $\hat {\cal E}$ is antisymmetric under 
$Q\leftrightarrow q$, the complete   
(Liouville) operator on the right-hand side of Eq.\,(\ref{Schroed}) has a symmetric spectrum 
with respect to zero and, in general, will not be bounded below.\\ \noindent   
%This {\it Kaplan-Sundrum energy-parity symmetry} has been invoked before to protect   
%a (near) zero cosmological constant which, otherwise, is threatened by 
%many orders of magnitude too large zeropoint energies~\cite{KS,Elze07}. 
\item Finally, the following observation will be important: 
\begin{equation}\label{Ezero}
\hat {\cal E}\equiv 0\;\;\Longleftrightarrow\;\;
\mbox{true potential}\;v(x)\;\mbox{is constant, linear, or harmonic} 
\;. \end{equation}  
The analogous vanishing of $\hat {\cal E}$ in a field theory is equivalent   
with having massive or massless free fields, with or without external sources, 
and with or without bilinear couplings. Generally, in all these cases, anharmonic 
forces or interactions are absent.    
\end{itemize} 
We emphasize that the coupling 
of the Hilbert space and its dual and the related lacking of a stable ground state, 
in general, show that our reformulation of Hamiltonian dynamics does not 
qualify as a proper quantum theory.  
However, we will motivate certain modifications from which dynamical aspects of     
quantum mechanics seem to emerge after all.  

%%%%%%%%%%%%%%%%%%%%%%%%%%%%%%%%%%%%%%%%%%%%%%%%%%%%%%%%%%%%%%%%%%% 
\section{Spatiotemporal discreteness and fluctuations it may cause} 
Here we present a simple-minded argument indicating that the discreteness 
of spacetime may be relevant for the emergence of quantum mechanical 
phenomena, turning the Liouville equation into the von\,Neumann~-~Lindblad equation, 
in particular. 

Given that spacetime is discrete, there must be a characteristic 
length scale, the Planck length, where the continuum description of all phenomena breaks down. 
This implies that one might overlook important traces of 
this atomistic structure by employing continuum quantities which allow to arbitrarily 
extrapolate their functional dependence to scales $l\approx l_{Pl}$, 
where $l_{Pl}$ denotes the Planck length. 

Instead of having a {\it coarse-grained potential} $V$, for 
example, the {\it true potential} $v$ should become piecewise defined 
somehow, when approaching smaller and smaller scales in the continuum picture. 
Thus, the function $V$ is an approximation to $v$ and the 
difference between the two must give rise to local 
{\it fluctuations} $\delta V$. Therefore, we set: 
\begin{equation}\label{Ansatz} 
v(\mathbf{x})=V(\mathbf{x})+\delta V(\mathbf{x}) 
\;\;, \end{equation} 
in order to relate the short distance behaviour   
to its coarse-grained description. 

Two sources of fluctuations can enter here: 
First, the spatiotemporal discreteness \cite{Sorkin07}. 
Second, the possibly discrete nature of interactions or matter.   
The latter might go beyond what is usually seen as {\it effects} of quantum 
mechanics. This has not been explored in parallel with the 
causal set theory of the deep structure of spacetime and will not 
be further discussed at present. 

%%%%%%%%%%%%%%%%%%%%%%%%%%%%%%%%%%%%%%%%%%%%%%%%%%%%%%%%%%%%%%%%%%%%%%%%%%%%%
\subsection{``Asymptotic freedom'' on causal sets, piecewise linear potentials 
and emergence of the von\,Neumann equation} 
However, there is an important ``asymptotic freedom'' effect, caused by  
spatiotemporal discreteness: on a (background) causal set, 
cross sections must fall to zero when the center of mass energy of 
two scattering particles reaches the Planck 
scale \cite{Sorkin09}.~\footnote{I thank Rafael Sorkin for the following illuminating 
description of this ``No Interaction Theorem'': {\it Imagine each particle or 
wave-packet as a ``world tube''
in Minkowski space. In the center of mass frame, the two tubes
make an ``X''. The center of the X, where the tubes meet, is the
interaction region. Now boost each particle to a very high
energy. Because of the Lorentz contraction, the tubes become so
flattened that the interaction region shrinks down to less than a
Planck volume. Hence there is (very likely) no element of the
causet to represent this interaction region, whence there is no
interaction.}} This is based on viewing the causal set  
{\it as if} generated by a Poisson process, ``sprinkling'' set elements with 
an average density of one element per Planckian volume, $(l_{Pl})^4$ --  
for example, into Minkowski space, if this is considered as the continuum limit. 

The ``asymptotic freedom'' effect suggests to consider the {\it true potential} function 
$v$ as {\it piecewise linear}, with the pieces -- simplices in higher dimensions -- 
characterized by short ``linearity lengths'' $\delta_n$. They must be, however, sufficiently larger than the Planck length, $\delta_n>l_{Pl}$, such that the continuum description 
is meaningful.~\footnote{Such considerations may deviate 
from the strategy to adapt continuum quantities to a discrete spacetime -- for example, 
by approximating a quartic potential by a quartic expression with support on the 
elements of a causal set. It seems a valuable alternative to look for a model of matter 
degrees of freedom and their interactions which is formulated in terms  
innate to a causal set, as far as possible. For free particles, work in this direction 
has been reported in Ref.~\cite{Philpottetal08}.}

With these remarks in mind, we reconsider the force term $\propto v'$ 
in Eq.\,(\ref{Schroed}). We rewrite this equation, admitting more than 
one spatial dimension, more simply as: 
\begin{equation}\label{SchroedR} 
i\partial_tf(\mathbf{Q},\mathbf{q};t)=[\hat H_0,\hat f]_{(\mathbf{Q},\mathbf{q};t)}
+(\mathbf{Q}-\mathbf{q})\!\cdot\!\mathbf{\nabla}
v(\frac{\mathbf{Q}+\mathbf{q}}{2})f(\mathbf{Q},\mathbf{q};t)
\;\;, \end{equation} 
where $\hat H_0 :=-\frac{1}{2}\Delta$, with $\Delta$ denoting the Laplacian, in an obvious generalization.    
Now, the term involving the derivative of the potential 
is related to the {\it would-be quantum mechanical} term -- the potential difference 
between points $Q$ and $q$ --  
as a linear mid-point approximation to an integral: 
\begin{equation}\label{approxInt} 
(\mathbf{Q}-\mathbf{q})\!\cdot\!\mathbf{\nabla} 
v(\frac{\mathbf{Q}+\mathbf{q}}{2})\;\approx\;\int_{\mathbf{q}}^{\mathbf{Q}} 
\mbox{d}\mathbf{s}\!\cdot\!\mathbf{\nabla} v(\mathbf{s})=v(\mathbf{Q})-v(\mathbf{q}) 
\;\;. \end{equation} 
The approximate relation becomes an 
exact equality in the cases covered by observation (\ref{Ezero}).   
  
More generally, the integral -- which yields {\it exactly} the quantum mechanical term -- can 
be decomposed into a sum of integrals along straight line segments: 
\begin{equation}\label{QMintegral} 
v(\mathbf{Q})-v(\mathbf{q})=\sum_n\int_{\mathbf{s_n}}^{\mathbf{s_{n+1}}} 
\mbox{d}\mathbf{s}\!\cdot\!\mathbf{\nabla} v(\mathbf{s})
=\sum_n(\mathbf{s_{n+1}}-\mathbf{s_n})\!\cdot\!\mathbf{\nabla} 
v(\frac{\mathbf{s_n}+\mathbf{s_{n+1}}}{2})
\;\;, \end{equation}
determined by a set of positions $\{ \mathbf{s_n}\}$ which match the 
piecewise linearity of the true potential
with the path of integration from $\mathbf{q}$ to $\mathbf{Q}$, i.e., with the  
linearity lengths, $\delta_n=|\mathbf{s_{n+1}}-\mathbf{s_{n}}|$. 

We conclude that within small volumes, the linear size of which is determined by 
the linearity length $\delta$ of the potential there, the Liouville equation 
of {\it classical} statistical mechanics, in the form of Eq.\,(\ref{Schroed})  
or Eq.\,(\ref{SchroedR}), is {\it indistinguishable} from the corresponding {\it quantum 
mechanical} von\,Neumann equation, which both refer to the continuum description.   

Furthermore, we learn from Eqs.\,(\ref{SchroedR})--(\ref{QMintegral}) 
that for larger distances, covering  
several linearity lengths $\delta_n$, the potential (commutator) term of the  
von\,Neumann operator is obtained as a {\it sum of classical contributions}. 
This summation is natural in view of the fact that each term adds an appropriate 
amount of energy to the potential difference $v(\mathbf{Q})-v(\mathbf{q})$, with one contribution 
per interval over which the potential is linear.  

In this way, we obtain the Hamilton operator  
for arbitrarily extended, piecewise linear potentials: 
\begin{equation}\label{Hv} 
\hat H=-\frac{1}{2}\partial_{\mathbf{x}}^{\;2}+v(\mathbf{x}) 
\;\;, \end{equation} 
in the coordinate representation, and readily  
infer the von\,Neumann equation in operator form: 
\begin{equation}\label{vNplp}
\partial_t\hat f=-i[\hat H,\hat f]
\;\;. \end{equation}
However, following the argument from Eq.\,(\ref{SchroedR}) to Eq.\,(\ref{vNplp}), 
there is a subtlety behind the last two equations here, concerning 
the composition of the simplices. 

In the continuum picture, each simplex forms a tiny ``box'' with 
a {\it sharply localized} boundary condition, which prohibits the leaking of probability 
to the outside -- as long as they are sufficiently separated. Bringing two such boxes    
next to each other, such that a common border region arises which is less than 
a Planck length ``thick'', the confining boundary condition will dissolve into a free 
boundary condition, no matter how it originated in the first place. This is another 
manifestation of the ``asymptotic freedom'' property, caused by the discrete 
spacetime structure. Thus, we imagine that the range of variables entering the 
matrix elements of $\hat f$ can adiabatically increase to the full range over 
both simplexes. 

While this ``Gedankenexperiment'' might seem plausible, a serious justification 
of the extension of the range of validity of Eq.\,(\ref{vNplp}) over 
more than one linearity length seems necessary. 
We intend to come back to this important point elsewhere.   

%%%%%%%%%%%%%%%%%%%%%%%%%%%%%%%%%%%%%%%%%%%%%%%%%%%%%%%%%%%%%%%%%%%%
\subsection{Coarse-graining, decoherence and localization mechanism}

In order to extend our considerations further to reach the scales 
where quantum mechanics is known to work, we have to address also the 
coarse-graining that must be involved. This cannot be analyzed  
rigorously without understanding how the common forces / interactions that we  
encounter relate to phenomena at much shorter distance scales, such as the 
linearity scale. 

However, qualitatively, we expect the coordinates to {\it fluctuate}, which enter the continuum 
description, in particular, when we write down the von\,Neumann equation with the 
potential terms obtained in Eq.\,(\ref{QMintegral}). -- Indeed, let us consider, 
for example, an event determined in the continuum picture. The position ascribed to 
it in a fixed Lorentz frame cannot be exactly maintained with time, 
if Minkowski space is a {\it continuum approximation} to an underlying discrete structure: 
to the future of this event there likely will be Planck scale size (or larger) patches 
of Minkowski space which contain no element of the approximated causal set. It 
appears to us that the event, respectively its successors, must change position by 
small amounts with time (``swerve''), effectively 
performing a random walk within its future lightcone \cite{Philpottetal08}. 

We can get a feeling for the linear size 
$\triangle r$ of these swerves by estimating the probability 
to have a spherical void of volume $\propto\triangle r^3$ that lasts for  
a Planck time interval. For the Poisson process mentioned earlier and with 
$\triangle r\gg l_{Pl}$, we find this to be  $e^{-(\triangle r/l_{Pl})^3}$. 
Therefore, uncertainties 
$\triangle r$ of the order of a typical distance pertaining to the Standard Model, 
for example, are extremely improbable. Nevertheless,  
at the linearity scale they may play an essential role in washing out the piecewise linear 
potentials.~\footnote{Similarly, there, they will have an effect on the kinetic 
terms of the Liouville or von\,Neumann operator; we simply take them as they are, since  
they yield the appropriate quantum mechanical operator. However, a detailed study of 
the emergence of such terms from a causal set perspective, of the wave operator in particular, 
has recently been performed \cite{Sorkin07}, eventually to be combined with 
considerations along the lines presented here.} 

By experience, smooth polynomial interactions 
play an essential role, say in the Standard Model and, to be derived from there, 
in phenomenological forces of physics at still lower energies. Since details of the 
coarse-graining are unknown, in particular those leading to highly 
constrained gauge symmetries, we now consider the phenomenological 
{\it Ansatz} of Eq.\,(\ref{Ansatz}), which connects {\it short distance 
behaviour at the linearity scale} with its {\it coarse-grained description}.  
In particular, this replaces a true, piecewise linear potential $v$ by a coarse-grained 
potential $V$ plus local fluctuations $\delta V$.    

The fluctuations are treated as white noise with mean and correlation, respectively, 
given by:  
\begin{eqnarray}\label{mean} 
\langle\delta V(\mathbf{x})\rangle &=&0 
\;\; , \\ [1ex] \label{correl} 
\langle\delta V(\mathbf{x})\delta V(\mathbf{y})\rangle 
&=&\nu^2(\mathbf{x})\delta (\mathbf{x}-\mathbf{y})/\delta (0)   
\;\;, \end{eqnarray} 
where $\nu (\mathbf{x})$ describes the width of the local distribution of 
fluctuations. This distribution cannot be assessed without further insight into   
how forces arise at small spacetime scales. 

We now implement Eq.\,(\ref{Ansatz}), $v(\mathbf{x})=V(\mathbf{x})+\delta V(\mathbf{x})$,  
in the von\,Neumann equation following from Eqs.\,(\ref{SchroedR})--(\ref{QMintegral}), 
with piecewise linear potential. This 
yields the von\,Neumann equation with the coarse-grained potential $V$ and additional  
fluctuating terms:   
\begin{equation}\label{vNL}    
\partial_t\hat f=-i[\hat H_{eff},\hat f]
\;\;, \end{equation}
where the effective Hamilton operator is now given by: 
\begin{eqnarray}\label{Heff} 
\hat H_{eff}&:=&\hat H+\delta V
\\ [1ex] \label{Heff1}
&:=&-\frac{1}{2}\partial_{\mathbf{x}}^{\;2}+V(\mathbf{x})+\delta V(\mathbf{x})
\;\;, \end{eqnarray} 
in the coordinate representation, cf. Eq.\,(\ref{vNplp}).  

In order to evaluate the influence of the fluctuations, we 
iterate Eq.\,(\ref{vNL}) once: 
\begin{equation}\label{vNL1} 
\partial_t^{\;2}\hat f=-i[\hat H,\partial_t\hat f]+\mbox{O}(\delta V )
-[\delta V,[\delta V,\hat f]
\;\;. \end{equation}
Integrating once, with the initial condition: 
\begin{equation}\label{initial}  
\partial_t\hat f|_{t=0}=-i[\hat H,\hat f]_{t=0}+\mbox{O}(\delta V )
\;\;, \end{equation} 
and averaging over the fluctuations, with the help of Eqs.\,(\ref{mean})--(\ref{correl}),
leads to:
\begin{equation}\label{vNL2} 
\partial_t\hat f=-i[\hat H,\hat f]
-\int_0^t\mbox{d}t'(\{\hat\nu^2,\hat f\}-2\hat\nu\hat f\hat\nu )
\;\;, \end{equation}
with the matrix elements 
$\langle\mathbf{x}|\{\hat\nu^2,\hat f\} |\mathbf{y}\rangle 
:=\nu^2(\mathbf{x})f(\mathbf{x},\mathbf{y})+f(\mathbf{x},\mathbf{y})\nu^2(\mathbf{y})$ 
and $\langle\mathbf{x}|\hat\nu\hat f\hat\nu |\mathbf{y}\rangle 
:=\nu^2(\mathbf{x})f(\mathbf{x},\mathbf{y})\delta (\mathbf{x}-\mathbf{y})/\delta (0)$.   
For sufficiently short times and slowly varying $\hat f$, we may use the 
approximation: 
\begin{equation}\label{vNLM} 
\partial_t\hat f\approx -i[\hat H,\hat f]
-t(\{\hat\nu^2,\hat f\}-2\hat\nu\hat f\hat\nu )
\;\;, \end{equation}
Thus, we arrive at a Markovian master equation with a dissipative {\it Lindblad term},  
in addition to the leading commutator, which is responsible for unitary 
quantum evolution \`a la von\,Neumann in the absence of dissipation.  

Linearity in the density matrix is an important feature of our resulting  
equations. In particular, in the form of Eqs.\,(\ref{vNL2})--(\ref{vNLM}),  
%It implies that the generator of time evolution is related to a contraction semigroup.  
the master equation preserves the normalization 
of $\hat f$, say $\mbox{Tr}\hat f=1$, which expresses the conservation of probability.  

It seems worth while to point out that we arrive at the standard, i.e., in 
particular {\it local}, quantum mechanical evolution equation -- even if modified 
by a less standard yet wellcome Lindblad term. 
However, we emphasize that the relation to the primordial deterministic degrees of freedom, 
which involves a Fourier transformation (cf. Section\,2), is highly nonlocal. 
This may ease the tension created by Bell's theorem, when it comes to deterministic 
(``hidden'') variables. 

The Lindblad term in Eq.\,(\ref{vNLM}) implies an interesting    
{\it decoherence and continuous localization mechanism}, which causes    
the decay of spatial superpositions (``Schr\"odinger cat states''). 
While the diagonal matrix elements of $\hat f$ are not affected by the Lindblad term, 
the off-diagonal matrix elements decay: 
\begin{equation}\label{decay}
f(\mathbf{x},\mathbf{y};t)
=f(\mathbf{x},\mathbf{y};0)\mbox{e}^{-\frac{1}{2}t^2(\nu^2(\mathbf{x})+\nu^2(\mathbf{y}))}
\;\;, \end{equation} 
where we neglected the effect of $\hat H$, for simplicity, which cannot 
stop the decay. Ultimately, spacetime discreteness produces this 
mechanism -- via the induced fluctuations in interactions. 

Since there is no theory yet to tell us about the correlation function $\nu^2$, 
it may suffice to point out that considerations of stochastic effects on  
quantum mechanics -- as it is -- have quite a history, see, for example,  
Refs.~\cite{GianCarlo09,BassiGhirardiRev,Diosi84,DiosiRev05} 
and the literature cited there. 
This is related to attempts to solve the measurement problem 
and to account for the apparent absence of spatial superposition states of macroscopic 
objects \cite{Penrose98,Adler03}. Also quantum gravity is conjectured to 
lead to characteristic stochastic effects \cite{Mavromatosetal92,Pullinetal08,Hu09}. 
In this context, various proposals 
for the analogue of our Lindblad operator $\hat\nu$ in Eq.\,(\ref{vNLM}) 
have been made as well and it has similarly been concluded that spatial 
superposition states decohere and decay. 

While these issues are not settled, our heuristic 
arguments suggest that  
if quantum mechanical behaviour indeed emerges dynamically, 
then the resolution of the measurement problem may be linked to this.   
`Prequantum' dynamics may account for 
an objective selection mechanism in accordance with the observed 
wave function collapse in measurement outcomes.  

Generally, Lindblad master equations present a large 
class of linear Markovian master equations, which are usually derived to describe open  
quantum systems that interact with particular environments~\cite{Diosi,Lindblad,Gorini}. 
Here spacetime itself forms the universal environment that ``measures'' all physical objects. 

We have derived this {\it quantum mechanical master equation}, beginning with {\it classical  statistical mechanics}, by incorporating several assumptions about the atomistic nature 
of spacetime, as well as about the nature of forces acting on matter. 

%%%%%%%%%%%%%%%%%%%%%
\section{Conclusions}

We have presented a heuristic discussion to the effect that quantum mechanics 
 -- together with a natural decoherence and continuous localization mechanism --   
is a consequence of classical statistics and concerns dynamics with respect to   
an atomistic spacetime that is observed with low resolution, i.e., at large distance scales. 
The spatiotemporal discreteness results in permanent information loss affecting all matter, 
when described in the corse-grained way that is appropriate with distances much larger than the 
Planck scale. We were motivated here by the theory of causal sets, where such a picture  
of spacetime does not follow from a quantization of gravity but is assumed as 
primary feature. 

There are a number of topics for further study. Work is in progress making newly 
use of the formalism adopted here, of using Hilbert 
space operators to describe classical statistics, and going beyond 
the early suggestions~\cite{Koopman,vN}. 

Furthermore, as pointed out in 
Ref.\,\cite{Elze09}, the problem of negative 
probabilities needs to be resolved, which is met when carrying the notion of probability density 
from the classical phase space theory over to the emergent quantum mechanical 
one, as we do. 

The generalization of 
the ideas sketched here to field theories has to face the fundamental problem 
to understand the nature of matter fields and the emergence of (the symmetries of) their interactions with  
decreasing spatiotemporal resolution. Our approach, paying special attention to 
piecewise linear potentials at sufficiently small scales, is in accordance 
with the suggestion of noninteracting fields at the Planck scale. There,   
the atomistic structure of spacetime becomes effective and -- for an underlying 
causal set -- induces such an ``asymptotic freedom property". 

\ack   
I am grateful to Lajos Di\'osi, Jose Isidro, Andrei Khrennikov and Rafael Sorkin for 
discussions or correspondence.

%%%%%%%%%%%%%%%%%%%%%%%%%%%

\end{document}